\documentclass[pre, preprint, a4paper]{revtex4-1}
\usepackage{graphicx}
\usepackage{amsmath}
\usepackage{amssymb}
\usepackage{hyperref}

\newcommand{\tr}{\operatorname{tr}}
\newcommand{\pd}[1]{\partial_{#1}}

\begin{document}

\title{Shock Wave Evolution into Strain Solitary Wave in Nonlinearly Elastic Solid Bar}

\author{F.E. Garbuzov}
\author{A.V. Belashov}
\author{A.A. Zhikhoreva}
\author{Y.M. Beltukov}
\author{I.V. Semenova}
\email{irina.semenova@mail.ioffe.ru}

\affiliation{Ioffe Institute, 26, Polytekhnicheskaya, St.Petersburg, 194021, Russia}



\begin{abstract}
In this paper we present thorough experimental observation of the process of shock wave transformation into a bulk strain solitary wave in a nonlinearly elastic solid bar made of polystyrene. A theoretical model based on the describing propagation of a plane elastic wave in a bar is developed with account made for material nonlinearity and viscosity. Numerical modeling performed on the base of the developed model with and without regard to viscosity demonstrated formation of a long stable disturbance at proper account made for viscoelastic properties of the bar material.
\end{abstract}

\maketitle

\section{Introduction}

In the last decades research on solitary waves became one of rapidly developing areas of mathematics and physics. Solitons were shown to exist in a wide variety of processes in optics, atomic physics, plasma physics, biophysics, fluid and solid mechanics, etc. \cite{Scott2005,Kharif2009,Boechler2010,Kevrekidis2008,Peyrard2004}.
In general a soliton is a localized wave that is formed due to the balance of nonlinear and dispersive effects and propagates with constant shape and speed.

Investigation of strain solitons propagating in the bulk of a solid body  (referred to as an elastic waveguide or simply a waveguide throughout this paper) is on the one hand important in view of potential unexpected generation of such waves in constructional elements that can cause their damage and on the other hand is promising for application in nondestructive testing. 
Strictly speaking, strain soliton can be formed only in an absolutely elastic solid where mechanical wave energy is conserved, while real solid materials, especially polymers, exhibit viscous properties which lead to energy dissipation and attenuation of strain waves. However, under certain conditions (e.g. when the viscosity is small) strain solitary waves can demonstrate a soliton-like behavior which makes them an important object of study.

Bulk strain solitary wave is a long trough-shaped elastic wave which is formed under favorable conditions from a short and powerful impact, usually in the form of a shock wave.
Unlike linear or shock waves, this wave does not exhibit significant decay in homogeneous waveguides and can propagate for much longer distances almost without shape transformation. 
Although major characteristics and behavior of formed strain solitary waves in different waveguides were studied sufficiently well (see e.g. \cite{TP2008,jap2010,jap2012,APL2014} and references therein), the details of shock wave transformation into a long solitary wave are not completely clear yet. This is partially due to significant difference in the initial and final wave shapes and amplitudes, their rapid change and a number of physical processes involved that complicates the experimental investigation of the solitary wave formation process. 

Optical techniques, in particular holographic interferometry and digital holography, proved to be quite suitable and informative in analysis of long bulk elastic waves in transparent waveguides (e.g. \cite{TP2008,WaMot2017,ApplSci2022}). While classical holographic interferometry with recording on high-resolution photographic materials required utilization of pulsed lasers and optical reconstruction of holograms, recent progress in fast-response global-shutter-assisted digital cameras allowed for simplification of the experimental procedure and usage of continuous-wave (CW) lasers for recording of wave patterns and computer-based hologram reconstruction. Therefore, processing of recorded digital holograms and data retrieval required  development of a number of specific numerical algorithms, in particular those of phase unwrapping \cite{lee2014single, goldstein1988satellite, wang2019one} and filtering \cite {uzan2013speckle, shevkunov2020hyperspectral, katkovnik2015sparse}. On the other hand such short powerful disturbances as shock waves are scarcely resolvable by holographic techniques that allow mostly for visualization of wave patterns \cite{Takayama2019}.  

Nonlinear elastic properties of the waveguide material play a key role in solitary wave formation and evolution. However, viscoelastic properties turn to be quite essential as well. In this paper we demonstrate how the initial shock wave transforms into the long solitary wave in the waveguide taking into account the viscoelastic effects, which suppress the formation of a number of short waves. 

The paper is organized as follows. First we describe the experimental procedure applied for generation and visualization of the process of solitary wave formation in a polymer waveguide.
Then we present experimental results showing evolution of the wave pattern while travelling along the waveguide. In further sections we describe the developed three-dimensional model of nonlinear viscoelastic solid and its simplification for a one-dimensional model. Numerical modeling of the process of shock wave evolution into the solitary wave is presented in the next section. And we finalize the paper with discussion of the results obtained and conclusions.

\section{Experimental procedure}

We have previously shown that strain solitary waves can be formed in waveguides made of glassy polymers (polystyrene, polymethylmethacrylate and polycarbonate) from an initial shock wave produced in water in the vicinity of the waveguide input (see \cite{TP2008,TPL2011} and references therein). 
In current experiments the shock wave was produced by explosive evaporation of the metallic foil by a pulsed Nd:YAG laser Spitlight 600 (Innolas, Germany) with pulse duration of 7 ns and pulse energy of 0.4 J at the wavelength of 532 nm.
The shock wave evolution into the strain solitary wave was analyzed in a bar-shaped waveguide, $10\times10$ mm in cross-section, 600 mm long made of transparent polystyrene (PS). 
Observation and monitoring of wave patterns in the bulk of the waveguide was performed by digital holographic technique based on registration of 2D distributions of phase shift of the recording wave front induced by variations of waveguide thickness and density formed by the waves under study. 

\begin{figure}[h!]
	\centering
	\includegraphics[width=15cm]{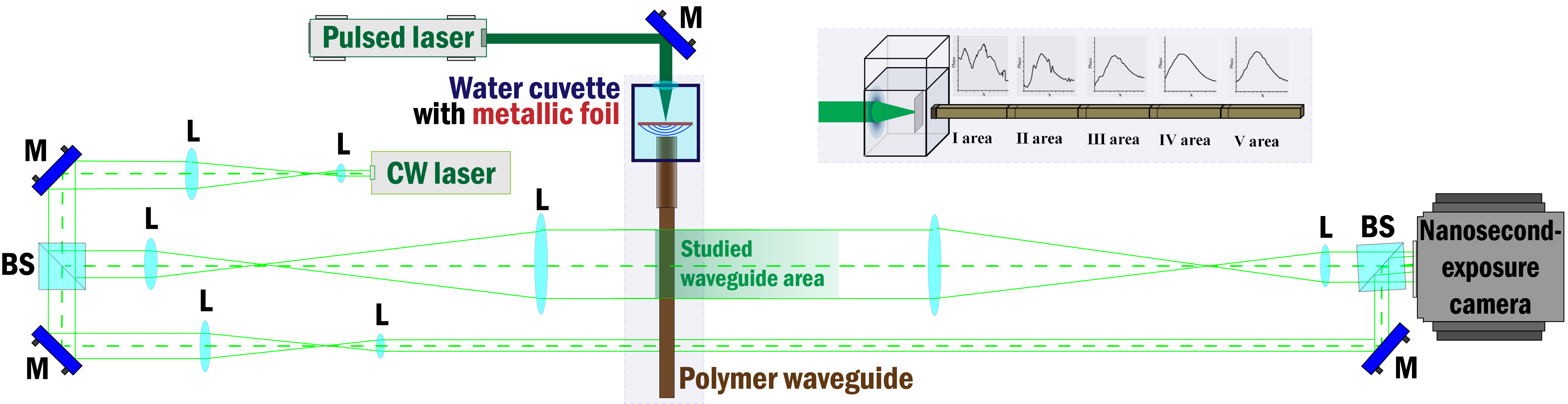}
	\caption{Optical schematic of the setup used for strain wave detection by means of digital holography. L are lenses, M are mirrors, BS are beamsplitters. Schematic of waveguide areas is shown in the inset.}
	\label{FigSetup}
\end{figure}

Holograms were recorded by a fast electro-optical camera Nanogate 24 (Nanoscan, Russia) with adjustable gain and exposure down to 10 ns. Synchronization of the laser pulse and camera shutter was provided by AM300 Dual Arbitrary Generator (Rohde$\&$Schwarz).
The short exposure time provided by the recording camera allowed for sufficiently fast registration of interference patterns, so that the displacement of the elastic wave propagating in the waveguide during the exposure time did not exceed the effective pixel size with regard to the reduction factor of the optical system.
The exposure time in our experiments was within the range of 70 -- 120 nanoseconds and was slightly varied to obtain high-quality holograms. The optical schematic of the experimental setup 
is shown in Fig. \ref{FigSetup}. Radiation from a probe semiconductor CW laser emitting at 532 nm was spatially filtered and divided into reference and object beams; the object beam was then expanded by a telescopic system to a diameter of 65 mm. A translation stage allowed us to shift the cuvette with the waveguide in the direction perpendicular to the object beam and to record wave patterns in different areas of the waveguide as shown in the inset in Fig. \ref{FigSetup}. For the sake of direct comparison  the sequential shifted fields of view overlapped for about 15 mm, so that areas in Fig. \ref{FigSetup} are about 50 mm long. 
The typical width of interference fringes in the hologram was adjusted to cover 8 -- 12 pixels of the camera matrix, that provided optimal performance of the reconstruction procedure of recorded holograms using the least squares algorithm   \cite{liebling2004complex}.  

\begin{figure}[h!]
	\centering
	\includegraphics[width=\textwidth]{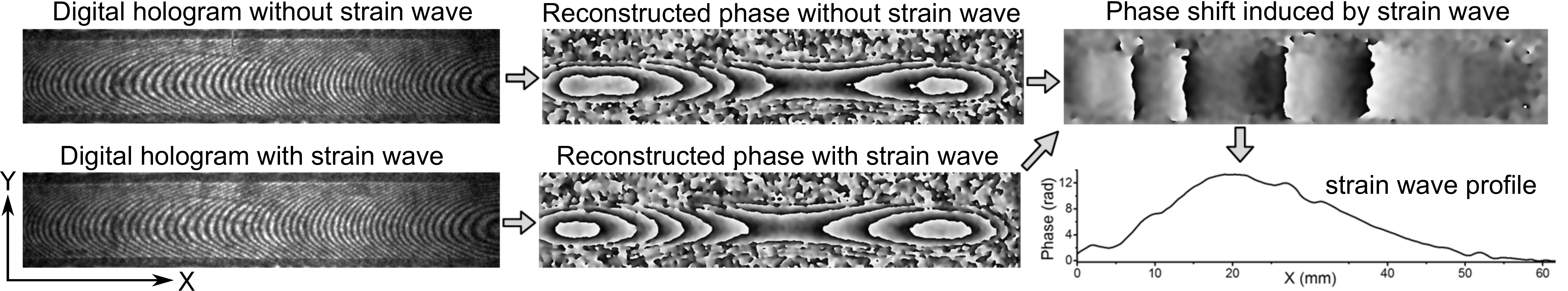}
	\caption{Procedure of strain wave reconstruction from recorded digital holograms.}
	\label{Figstrainwaverec}
\end{figure}

To compensate distortions caused by various factors including those due to non-ideal shape of the waveguide, the phase shift distribution introduced by the elastic waves into the probe wave front was determined as a difference between the reconstructed phase images of the waveguide in the presence and absence of the strain wave.
We note that the sidewalls of the polystyrene bar were slightly distorted and the probe wavefront that passed through it had a somewhat cylindrical shape, that is clearly seen in the examples of digital holograms shown in Fig. \ref{Figstrainwaverec}. 
Since strain waves of high amplitude can introduce a phase shift noticeably exceeding 2$\pi$ radian, the obtained phase distribution corresponding to the strain wave was unwrapped \cite{ghiglia1998two} (see Fig. \ref{Figstrainwaverec}). If a strain wave is uniform along the Y-axis, its one-dimensional profile can be found by data averaging along the Y-axis. Mention that in some cases it is impossible to reconstruct phase shift distributions in the vicinity of waveguide edges because of their curvature.

The solitary wave parameters can be obtained from phase distributions: the wave width is determined as the distance between zero levels of phase shift and wave amplitude can be calculated as \cite{pump-probe2018}:
\begin{equation}
    A = \frac{\phi\lambda_{\rm rec}}{2{\pi}h[(n-1)\nu+C(1-2\nu)]},
\end{equation}
where $\phi$ is the maximal phase shift, $\lambda_{\rm rec}$ is the recording light wavelength, $h$ is the bar thickness, $n$ and $\nu$ are, respectively, the refractive index and Poisson's coefficient of the bar material. The dimensionless coefficient C describes the dependence between the local density variations and refractive index: $\Delta n = C \Delta \rho/\rho$. According to the Lorenz-Lorentz equation C can be estimated for polystyrene as $C \approx n - 1$  with the precision of about
8\% \cite{vedam1976variation}.

Note that for recording of long complex wave patterns the available field of view may be insufficient. In these cases a synthetic large field of view is required for reliable observation of wave patterns. Such a synthetic field of view can be obtained by stitching together several wave profiles collected by reconstructing and processing of a set of  phase images of the wave pattern at sequential time moments (see \cite{pump-probe2018} for details). 

\section{Experimental results}

\begin{figure}[h!]
	\centering
	\includegraphics[width=15.5cm]{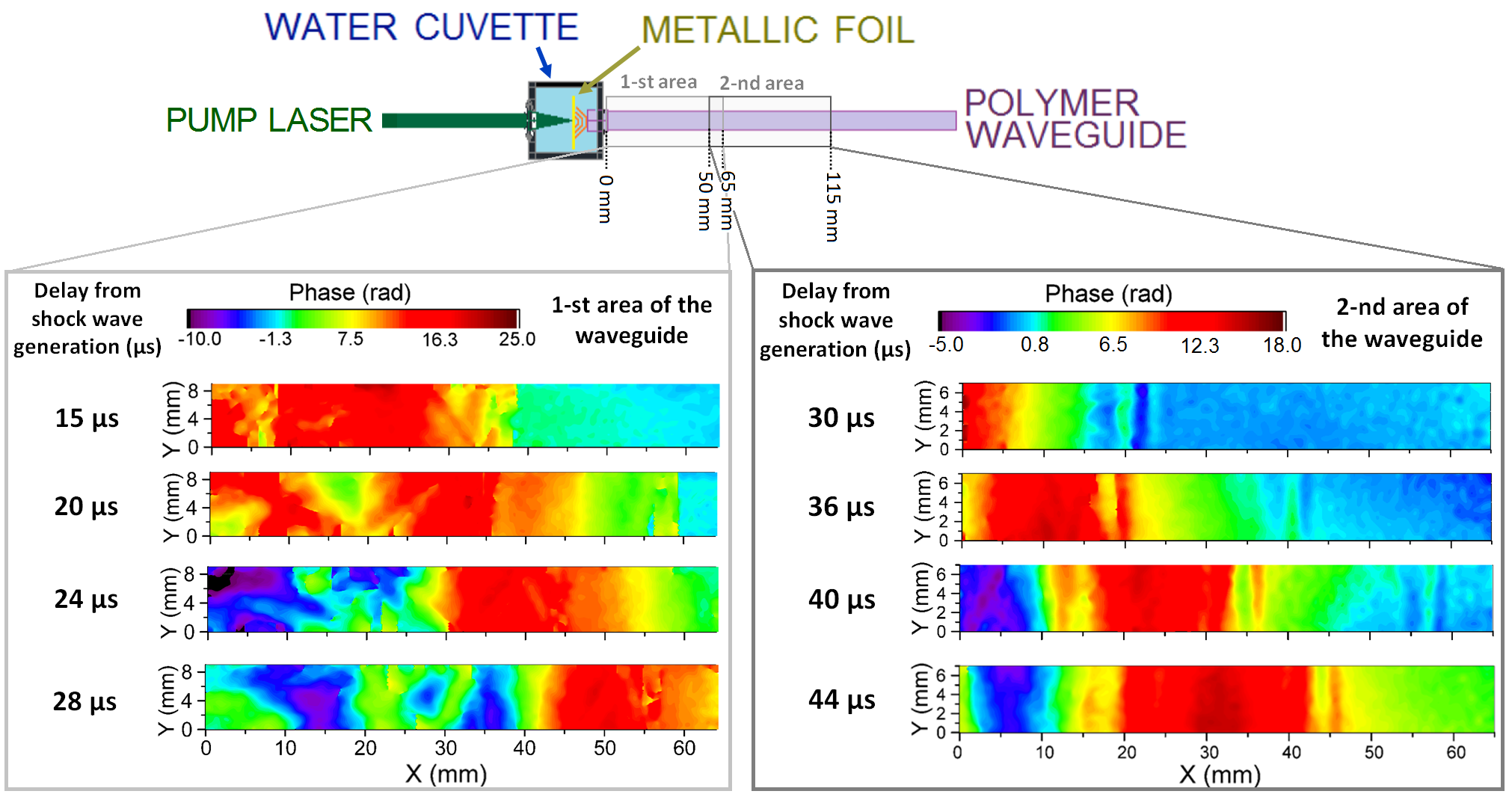}
	\caption{Phase distributions demonstrating wave patterns in the first (left) and second (right) areas of the polystyrene bar at different delays between the moments of shock wave generation and strain wave detection.}
	\label{FigEvolutionDH}
\end{figure}

\begin{figure}[h!]
	\centering
	\includegraphics[width=16cm]{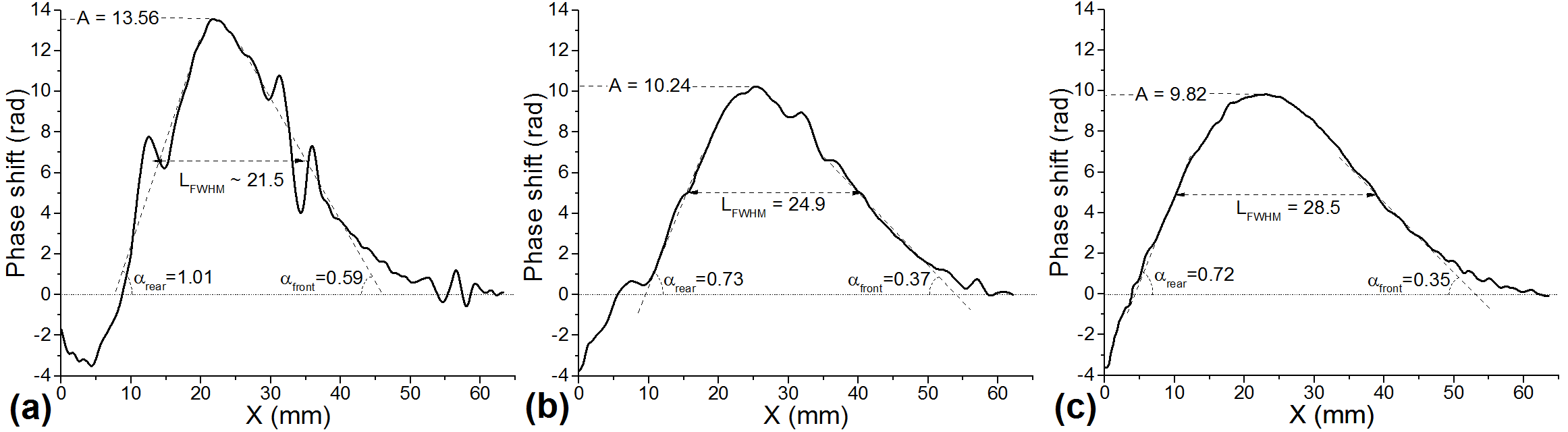}
	\caption{Y-averaged contours of the strain waves recorded in three areas of the polystyrene bar: (a) II, (b) III, (c) IV. Waves move from left to right. The coordinate X originates from the beginning of each measurement area.}
	\label{contours}
\end{figure}

We have previously shown \cite{TP2008} that a shock wave in water produced by laser evaporation of the metallic foil consists of a short ($\sim$ 0.1 \textmu m) powerful compression peak followed by a relatively long ($\sim$ 1 mm) rarefaction area. When entering the waveguide this wave produces a complex disturbance and relatively quickly looses its energy. 
However within the first couple of centimeters of the waveguide the initial shock wave is still too narrow to be resolved spatially and too powerful to be reconstructed in terms of phase shift of the recording wavefront (see \cite{TP2008,GKS2019} for details). 
Figure  \ref{FigEvolutionDH} presents a set of phase shift distributions characterizing evolution of the shock wave in the first two neighboring areas from the bar input (I and II areas as shown in Fig.~\ref{FigSetup}). In each area phase distributions were recorded at different delays between the laser pulse and camera shutter allowing for monitoring wave evolution within the field of view. As can be seen in Fig.~\ref{FigEvolutionDH} at the beginning of the waveguide, in the first area, the wave pattern is irregular representing a remainder of the initial shock wave followed by a long disturbance propagating with lower velocity. The wave pattern is nonuniform along the Y axis that does not allow for obtaining a reliable averaged contour. In the second area the wave pattern becomes more regular. The remainder of the shock wave outrun the general disturbance, it attenuated noticeably but is still visible. The long disturbance became more uniform, however it still has some relatively sharp peaks on its fronts. The plot of Y-averaged phase shift in this disturbance is shown in Fig. \ref{contours}a.  



\begin{figure}[h!]
	\centering
	\includegraphics[width=15.5cm]{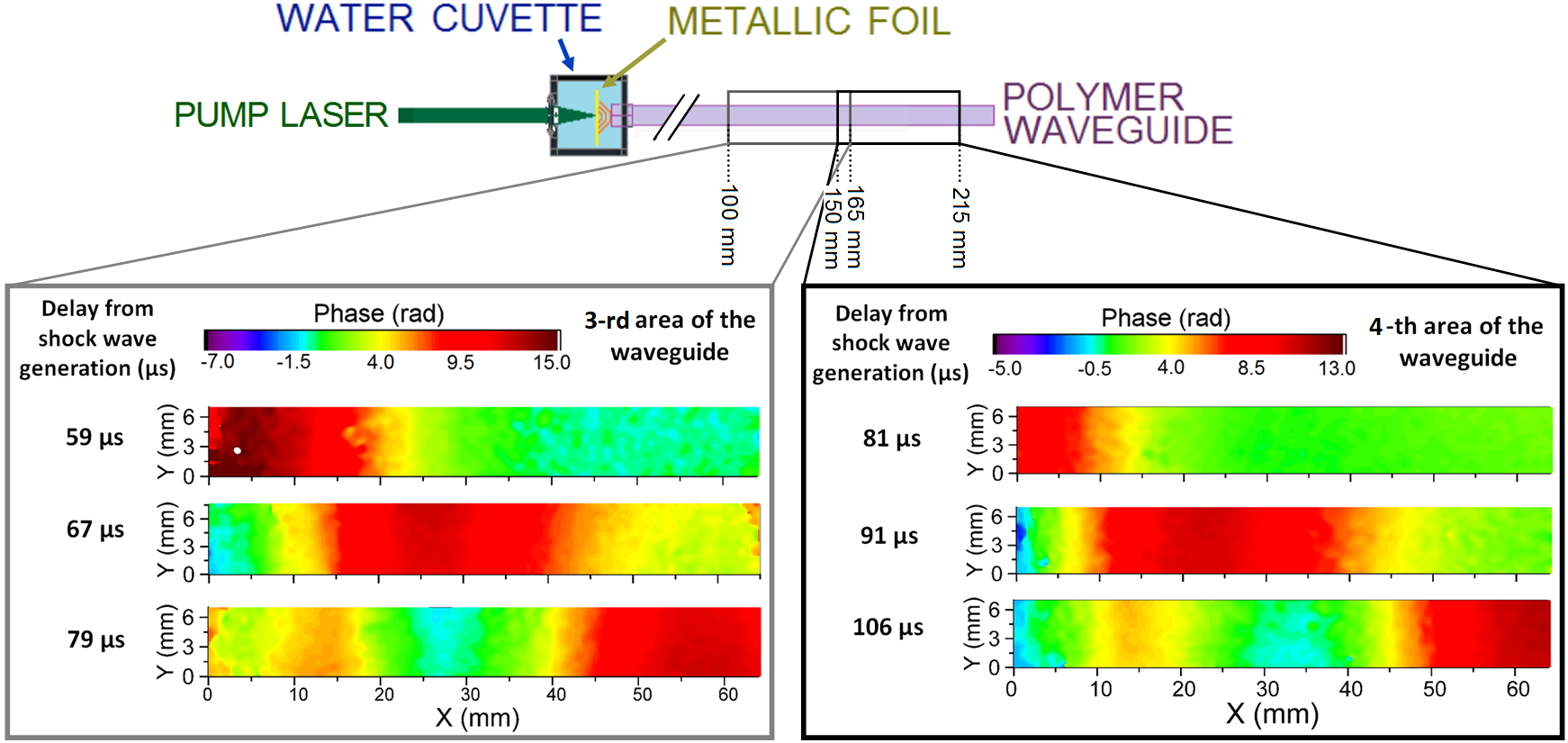}
	\caption{Phase distributions demonstrating wave patterns in the third (left) and fourth (right) areas of the polystyrene bar at different delays between the moments of shock wave generation and strain wave detection.}
	\label{IV-V-phase}
\end{figure}

Figure \ref{IV-V-phase} demonstrates sets of phase shift distributions characterizing wave patterns in the III-rd (100 -- 165 mm from the bar input) and IV-th (150 -- 215 mm) areas of the bar. The corresponding Y-averaged phase profiles are plotted in Fig. \ref{contours}.
As can be seen in Fig. \ref{IV-V-phase} the leading shock wave attenuated completely, the recorded phase shift distributions became smooth, with no abrupt changes, and sufficiently uniform along the Y axis, that allowed for estimation of the following wave parameters: amplitude, slopes of the leading and trailing edges, full width at half maximum $L_{\rm FWHM}$, width at the $1/e$ level $L_{\rm exp}$. The parameters of waves recorded in the I, II, III and IV areas are summarized in Table \ref{Tab1}. 

As can be seen from Table \ref{Tab1} while travelling along the waveguide the wave attenuated, its amplitude decreased and its width increased. At the same time the leading edge showed a tendency to be flatter with respect to the trailing edge since the ratio $\alpha_{\rm front}/\alpha_{\rm rear}$ gradually decreased with the propagation distance.
This behavior is confirmed by our previous experiments on monitoring of solitary wave evolution at longer distances, see \cite{TP2008} where we have demonstrated that the solitary wave still exists at the distances of several tens of centimeters and its attenuation coefficient of about 0.005 cm$^{-1}$ in polystyrene allows to expect attenuation in $e$ times at the distance of about 2 m.



\begin{table}
\centering
\caption{\label{Tab1}Parameters of strain waves recorded in the I-st, II-nd, III-rd and IV-th areas of the polystyrene bar.}
\begin{tabular}{c|ccccc}
Area & \textbf{$\phi_{\rm max}$} & \textbf{$\alpha_{\rm front}$} & \textbf{$\alpha_{\rm rear}$} & \textbf{$L_{\rm FWHM}$} &  \textbf{$L_{\rm exp}$}\\
 & (rad) & (rad/mm) & (rad/mm) & (mm) &  (mm)\\
\hline
I & 17.74 & 0.82 & & &  \\
II & 13.56$\pm$0.6 & 0.59$\pm$0.03 & 1.01$\pm$0.06 & 21.5$\pm$0.7 & \\
III & 10.24$\pm$0.4 &  0.37$\pm$0.02 & 0.73$\pm$0.04 & 24.9$\pm$0.5 & 29.2$\pm$0.7 \\
IV & 9.82$\pm$0.3 &  0.35$\pm$0.02 & 0.72$\pm$0.04 & 28.5$\pm$0.5 &  34.0$\pm$0.7 \\
\end{tabular}
\end{table}




\section{Three-dimensional model of nonlinear viscoelastic body}
Experiments showed the initial steps of solitary wave formation, i.e. the decay of a short strain wave generated by the shock wave in the water cuvette and formation of a long strain wave behind the short one. This process requires theoretical explanation which we start in this section by describing the full three-dimensional model of the bar. Polymeric materials, which the bar is made of, exhibit viscoelastic and nonlinear properties which we include in the model.

In our work we consider solid bodies which undergo small but finite strains. These strains are described by the Green-Lagrange finite strain tensor $\mathcal{E}$ with the nonlinear component which we refer to as geometric nonlinearity:
\begin{equation}\label{eq:strain}
    \mathcal{E}_{\alpha\beta} = \frac12 \left(\frac{\partial U_\alpha}{\partial r_\beta} + \frac{\partial U_\beta}{\partial r_\alpha} + \frac{\partial U_\gamma}{\partial r_\beta} \frac{\partial U_\gamma}{\partial r_\alpha} \right).
\end{equation}
Here, the Einstein notation is used for the Greek indices, $U_\alpha$ is the displacement along the axis $\alpha$ and $r$ denotes material coordinate vector (Lagrangian approach).

The elastic response of a body is defined by the potential energy density $\Pi$, which we take in the form of a truncated power series in strain components with all possible quadratic terms corresponding to the mechanically linear material and all cubic terms that are the next order corrections and which we refer to as physical nonlinearity:
\begin{equation}\label{eq:pot_en}
    \Pi = {\frac{\lambda+2\mu}{2}I_1^2(\mathcal{E}) - 2\mu I_2(\mathcal{E}) + \frac{l + 2m}{3}I_1^3(\mathcal{E}) - 2m I_1(\mathcal{E}) I_2(\mathcal{E}) + n I_3(\mathcal{E})}.
\end{equation}
Here $\lambda$ and $\mu$ are the Lam\'e (linear) elastic moduli, $l$, $m$ and $n$ are the Murnaghan (nonlinear) elastic moduli and $I_1(\mathcal{E}) = \tr\mathcal{E}$, $I_2(\mathcal{E}) = [(\tr\mathcal{E})^2 - \tr\mathcal{E}^2]/2$, and $I_3(\mathcal{E}) = \det \mathcal{E}$ denote invariants of the strain tensor~\cite{Samsonov2001}. This expression for the potential energy density has the most general form for the small but finite strains in an isotropic material. This model should not be confused with hyperelastic models (e.g., Neo-Hookean, Mooney-Rivlin, etc.) which are designed to describe large strains in rubber-like materials.

The elastodynamics of a body is defined by the stationary point of the action functional which is the time integral of the Lagrangian $L$: 
\begin{equation}\label{eq:lagr}
    L = {\int_\Omega \mathcal{L} dV - \int_{\partial\Omega}U_\alpha F_\alpha dS,}
\end{equation}
where $F$ is the external boundary stress, $\Omega$ is the volume of the body, $\mathcal{L} = K - \Pi$ is the Lagrangian volume density, $K = \frac12\rho \dot U_\alpha \dot U_\alpha$ is the kinetic energy density and $\rho$ is the material density. The unknown displacements $U_\alpha$ have to satisfy the Euler-Lagrange equations for the action functional to be at its stationary point
\begin{equation}
    \frac{\partial}{\partial t} \frac{\partial\mathcal{L}}{\partial\dot U_\alpha} + \frac{\partial}{\partial r_\beta} \frac{\partial\mathcal{L}}{\partial(\partial_\beta U_\alpha)} - \frac{\partial\mathcal{L}}{\partial U_{\alpha}} = 0,
\end{equation}
where dot denotes the time derivative and $\partial_\beta$ denotes the derivative with respect to $r_\beta$. These equations yield the equations of motion which have to be complemented by the boundary conditions as follows:
\begin{gather}
    \label{eq:dyn}
    \rho \ddot{U}_\alpha = {\frac{\partial P_{\alpha\beta}}{\partial r_\beta}, \quad r\in\Omega}, \\
    \label{eq:bc}
    P_{\alpha\beta} n_\beta = F_\alpha, \quad r\in\partial\Omega,
\end{gather}
where $n$ is the surface normal vector and the first Piola-Kirchhoff stress tensor $P_{\alpha\beta}$ is defined as
\begin{equation}
    P_{\alpha\beta} = \frac{\partial\Pi}{\partial(\partial_{\beta}U_\alpha)} = \left(\delta_{\alpha\gamma} + \partial_{\gamma} U_\alpha\right) \left(S^{\rm lin}_{\gamma\beta} + S^{\rm nl}_{\gamma\beta}\right).  \label{eq:P}
\end{equation}
Here linear and nonlinear second Piola-Kirchhoff stress tensors are:
\begin{gather}
    S^{\rm lin}_{\alpha\beta} = \lambda \mathcal{E}_{\gamma\gamma}\delta_{\alpha\beta} + 2\mu \mathcal{E}_{\alpha\beta},  \label{eq:Slin} \\
    S^{\rm nl}_{\alpha\beta} = l \bigl(\mathcal{E}_{\gamma\gamma}\bigr)^2\delta_{\alpha\beta} + (2m-n) \bigl(\mathcal{E}_{\gamma\gamma} \mathcal{E}_{\alpha\beta} - I_2(\mathcal{E})\delta_{\alpha\beta}\bigr) + n \mathcal{E}_{\alpha\gamma}\mathcal{E}_{\gamma\beta}.   \label{eq:Snl}
\end{gather}
In the general case, stiffness is a retarded integral operator that expresses the influence of material history on its current state and the kernel of these operators determines viscoelastic properties of the material~(\cite{HowellMechanics}). The main viscoelastic effects can be taken into account in the linear stress tensor $\smash{S^{\rm lin}_{\alpha\beta}}$ using retarded integral operators $\hat{\lambda}$ and $\hat{\mu}$:
\begin{equation}
    S^{\rm lin}_{\alpha\beta} = \hat{\lambda} \mathcal{E}_{\gamma\gamma}\delta_{\alpha\beta} + 2\hat{\mu} \mathcal{E}_{\alpha\beta}. \label{eq:Slin_ve}
\end{equation}
We could also include the viscous effects in the nonlinear stress tensor $\smash{S^{\rm nl}_{\alpha\beta}}$, but it leads to unnecessary complication of the model and goes beyond the scope of the current work.

In our work we use the generalized Maxwell model with many characteristic relaxation times, which allows us to cover a wide range of frequencies where glassy polymers exhibit viscous properties. In this model operators $\hat{\lambda}$ and $\hat{\mu}$ act on an arbitrary function $f(t)$ at time $t$ in the following way:
\begin{subequations}
\label{eq:maxwell}
\begin{gather}
    \hat{\lambda} f(t) = \lambda f(t) + \hat\xi \dot{f}(t) = \lambda f(t) + \sum_{s=1}^{N} \xi_s \int_{-\infty}^{t} e^{-\frac{t-t'}{\tau_s}} \dot{f}(t') dt',    \label{eq:lambda_op}\\
    \hat{\mu} f(t) = \mu f(t) + \hat\eta \dot{f}(t) = \mu f(t) + \sum_{s=1}^{N} \eta_s \int_{-\infty}^{t} e^{-\frac{t-t'}{\tau_s}} \dot{f}(t') dt',  \label{eq:mu_op}
\end{gather}
\end{subequations}
where $N$ is the number of relaxation times $\tau_s$ with dilatational viscosity moduli $\xi_s$ and shear viscosity moduli $\eta_s$ (see~\cite{HowellMechanics} for details).

The system of equations (\ref{eq:dyn}) -- (\ref{eq:P}), (\ref{eq:Snl}), (\ref{eq:Slin_ve}) defines the dynamics of nonlinear viscoelastic body. For the generalized Maxwell model~\eqref{eq:maxwell}, the integro-differential equation (\ref{eq:Slin_ve}) can be written as a system of $N$ differential equations using the retarded strain rates $q^{(s)}_{\alpha\beta}$ as follows:
\begin{gather} 
    q_{\alpha\beta}^{(s)} + \tau_s \dot{q}_{\alpha\beta}^{(s)} = \mathcal{\dot E}_{\alpha\beta}, \quad s=1,\dots,N,\\
    S^{\rm lin}_{\alpha\beta} = \lambda \mathcal{E}_{\gamma\gamma} \delta_{\alpha\beta} + 2\mu \mathcal{E}_{\alpha\beta} + \sum_{s=1}^N\left[\xi_s q^{(s)}_{\gamma\gamma} \delta_{\alpha\beta} + 2\eta_s q_{\alpha\beta}^{(s)}\right]. \label{eq:damping}
\end{gather}

The presented theory is rather general and the equations shown here are probably impossible to solve exactly even for the bodies of simple geometry. However, these equations can either be simplified to account only for a particular wave processes, or they can be simulated numerically. We chose both paths which are described in the following sections.

\section{One-dimensional model for plane waves travelling along the bar}

\begin{figure}
	\centering
	\includegraphics[width=15cm]{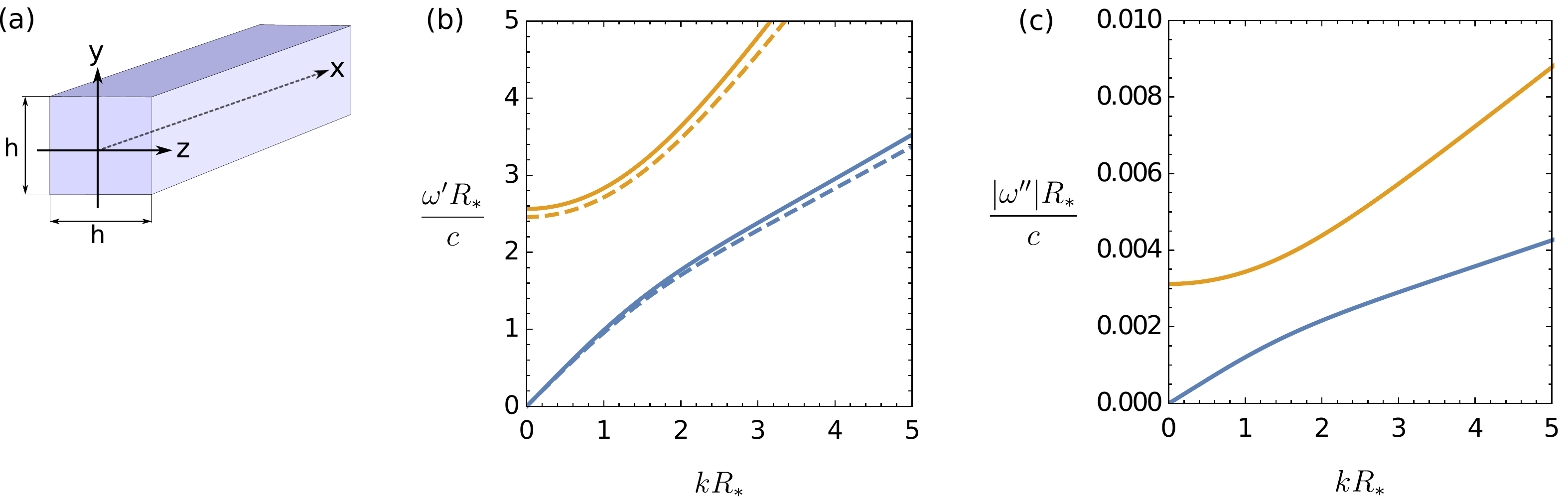}
	\caption{a) Polystyrene bar with the thickness $h$. b) Dispersion of linear waves in the one-dimensional model of polystyrene bar given as a dependence of the scaled frequency $\omega' R_*/c$ as a function of scaled wavenumber $kR_*$. Solid lines show the dispersion for viscoelastic model (\ref{eq:1d}) with parameters described in Section~\ref{sec:num}. Dashed lines show the dispersion of non-viscous model (\ref{eq:elastic1d}), which depends only on the Poisson ratio $\nu=0.38$ for the given scale. 
	c) The corresponding damping of linear waves in the one-dimensional viscoelastic model.}
	\label{fig:dispersion}
\end{figure}

We consider a symmetric bar of square cross-section of the thickness $h$ shown in Fig.~\ref{fig:dispersion}a. To derive a simplified model we apply several assumptions limiting the motion to plane waves propagating along the bar axis only. First, we assume that the bar cross-section remains flat and normal to the bar axis which makes the longitudinal displacement independent of the transverse coordinates $y$ and $z$:
\begin{equation}
    U_x(x,y,z,t) = u(x, t). \label{plane_sec}
\end{equation}
Second, we assume that the bar cross-section is uniformly deformed in $yz$ plane:
\begin{gather}
    U_y(x,y,z,t) = yv(x,t), \label{lin_transverse1}\\
    U_z(x,y,z,t) = zv(x,t), \label{lin_transverse2}
\end{gather}
where coordinates $y$ and $z$ originate from the center of mass of the cross-section. It can be shown that these assumptions are asymptotically satisfied both for long and short waves travelling along the bar axis. Here, we apply these assumptions beyond their proven range of applicability to derive a one-dimensional model for any wavelength. However, as we will show later, it turns out that the resulting simple model describes well the process of formation of the main long wave from the short one.

For assumptions~\eqref{plane_sec} --~\eqref{lin_transverse2} and non-viscous elastic moduli $\lambda$ and $\mu$,  one can obtain Lagrange equations using the Lagrangian density $\cal{L}$ integrated over $y$ and $z$, i.e. over the bar cross-section. Substituting the viscoelastic moduli $\hat\lambda$ and $\hat\mu$ into the resulting equations gives
\begin{subequations}
\label{eq:1d}
\begin{gather}
    \rho\ddot{u} = \left(\hat\lambda + 2\hat\mu\right) \pd{xx}u + 2 \hat \lambda \pd{x}v + \pd{x}f_{\rm nl}, \\
    \rho\ddot{v} = \hat \mu \pd{xx}v - \frac{2}{R_*^2}\left[ \hat\lambda \pd{x}u + 2\left(\hat\lambda+\hat\mu\right)v + g_{\rm nl}\right],
\end{gather}
\end{subequations}
where $R_*^2 = \langle y^2 + z^2\rangle$ is the mean radius squared taken over the cross-section. For the square bar under consideration $R_*^2=h^2/6$. The nonlinear terms $f_{\rm nl}$ and $g_{\rm nl}$ are given in Appendix~\ref{app}. For a finite bar these equations have to be complemented by the boundary conditions. If the bar has a flat edge at $x=0$ subjected to uniform pressure $F$ along $x$ direction, the corresponding boundary conditions have the following form 
\begin{subequations}
\label{eq:1d_bc}
\begin{align}
    &(\hat\lambda+2\hat\mu)\pd{x}u + 2\hat\lambda v + f_{\rm nl} = -F,\\
    &\pd{x}v = 0.
\end{align}
\end{subequations}

Equations~\eqref{eq:1d} constitute the coupled one-dimensional system of integro-differential equations ($\hat{\lambda}$ and $\hat{\mu}$ are the retarded integral operators) describing plane strain waves in a viscoelastic bar. This system is still hard to solve exactly, however the numerical simulation of it is way faster than for the full three-dimensional model. For the analytical study, different asymptotical regimes will be considered.

We start the analysis of the derived model~\eqref{eq:1d} in the linear regime, when all nonlinear terms can be neglected. The important special case of this regime is when the viscosity is also neglected and $\hat\lambda = \lambda$, $\hat\mu=\mu$. In the short-wave limit (wavelength is much smaller than $R_*$) one obtains two solutions of Eqs.~\eqref{eq:1d}: short longitudinal waves (P-waves) with the velocity $c_p = \sqrt{(\lambda+2\mu)/\rho}$ and short shear waves (S-waves) with the velocity $c_s = \sqrt{\mu/\rho}$. In the opposite limit of long waves one obtains two other solutions: long longitudinal waves with the velocity $c = \sqrt{E/\rho}$ defined by the Young modulus $E = \mu(3\lambda+2\mu)/(\lambda+\mu)$ and breathing mode with the frequency squared $\omega^2 = 4 (c_p^2 - c_s^2)/R_*^2$ and zero group velocity. General linear waves of an arbitrary wavelength are described by the equation 
\begin{equation}
    \pd{tt}u - c^2\pd{xx} u + \frac{R_*^2}{4\bigl(c_p^2 - c_s^2\bigr)} \Bigl(\pd{tttt}u - \bigl(c_p^2 + c_s^2\bigr)\pd{xxtt}u + c_p^2c_s^2\pd{xxxx}u\Bigr) = 0.   \label{eq:elastic1d}
\end{equation}
This equation has two dispersion curves, which are presented in Fig.~\ref{fig:dispersion}b by dashed lines and have the above-mentioned asymptotics. 

If viscosity is included but nonlinearity is still neglected the model describes linear elastic waves subjected to viscoelastic relaxation (damping). In this case one can introduce dynamic (complex) Lam\'e elastic moduli $\tilde{\lambda}(\omega) = \lambda'(\omega) + i \lambda''(\omega)$ and $\tilde{\mu}(\omega) = \mu'(\omega) + i \mu''(\omega)$ with real (storage) and imaginary (loss) components:
\begin{subequations}
\label{lam_mu_complex}
\begin{gather}
    \lambda'(\omega) = \lambda + \sum_s \frac{\xi_s\omega^2\tau_s^2}{1+\omega^2\tau_s^2}, \quad \lambda''(\omega) = \sum_s \frac{\xi_s\omega\tau_s}{1+\omega^2\tau_s^2}, \\
    \mu'(\omega) = \mu + \sum_s \frac{\eta_s\omega^2\tau_s^2}{1+\omega^2\tau_s^2}, \quad \mu''(\omega) = \sum_s \frac{\eta_s\omega\tau_s}{1+\omega^2\tau_s^2}. 
\end{gather}
\end{subequations}
Using complex elastic moduli $\tilde{\lambda}(\omega)$ and $\tilde{\mu}(\omega)$ one can find linear waves of the form $u,v\sim e^{ikx-i\omega t}$. For a given real wavevector $k$, one obtains complex frequency $\omega=\omega'+i\omega''$, where $\omega''$ describes the damping. Figure~\ref{fig:dispersion}c shows the damping of waves in polystyrene bar with realistic parameters, which will be used in the numerical modelling (see the next section for details). One can see that short waves with large wavenumbers are attenuated much faster than long longitudinal waves (lower curve for small wavenumbers). The breathing mode (upper curve for small wavenumbers) has relatively large damping and almost zero group velocity (Fig.~\ref{fig:dispersion}b,c).

The nonlinear case requires much more complicated analysis. However, the most important regime is the long-wave regime, when the solitary waves can be stabilized by the balance between the dispersion and the nonlinearity. 
In this case, when viscosity is neglected, we obtain the nonlinear equation of the Boussinesq type:
\begin{equation}
    \pd{tt}u - c^2\pd{xx} u + \frac{R_*^2}{4\bigl(c_p^2 - c_s^2\bigr)} \Bigl(\pd{tttt}u - \bigl(c_p^2 + c_s^2\bigr)\pd{xxtt}u + c_p^2c_s^2\pd{xxxx}u\Bigr) - \frac{\beta_s}{\rho}(\pd{x}u)(\pd{xx}u) = 0.  \label{eq:nonlinear}
\end{equation}
Here $\beta_s=3E+2l(1-2\nu)^3 + 4m(1+\nu)^2(1-2\nu) + 6n\nu^2$ is the nonlinearity modulus~\cite{Samsonov2001}, where $\nu=\lambda/(2\lambda + 2\mu)$ is the Poisson's ratio. Equation (\ref{eq:nonlinear}) has one-parameter family of exact solitary wave (soliton) solutions:
\begin{equation}
    \pd{x} u = A \operatorname{sech}^2 \frac{x - st}{L},
\end{equation}
where $s$, $A$, and $L$ are velocity, amplitude, and length of the soliton, respectively, with the following relation between them
\begin{equation}
    A = \frac{3E}{\beta_s}\left(\frac{s^2}{c^2}-1\right), \quad L^2 = \frac{\bigl(c_p^2-s^2\bigr)\bigl(s^2 - c_s^2\bigr)}{\bigl(c_p^2-c_s^2\bigr)\bigl(s^2 - c^2\bigr)}R_*^2.
\end{equation}
A more rigorous asymptotic analysis of the model can be carried out, assuming that the viscous terms are small. In this case, a soliton-like solution with slowly varying amplitude, velocity, and length could be obtained~\cite{Kivshar1989}. However, this analysis is beyond the scope of our work, since we are focused on the initial viscoelastic evolution of strain waves.

The derived equation~\eqref{eq:nonlinear} can be viewed as the standard equation for the longitudinal waves travelling at the speed $c$ with small corrections in the form of nonlinear and dispersive terms. This equation has exactly the same nonlinear term and asymptotically equivalent dispersive terms as in the previously obtained Boussinesq-type models for the long waves in rods and bars~\cite{KhusnSamsPhysRev2008, GKS2019}.

\section{Numerical modelling}
\label{sec:num}

To perform numerical modelling of polystyrene bar, material parameters should represent realistic viscoelastic properties of polystyrene. The density was taken as ${\rho = 1.06\text{ g/cm}^3}$. To simulate the propagation of elastic waves excited by a short impact, one should take into account various relaxation processes in the most important range of relaxation times, which lay between 0.01 \textmu s and 100 \textmu s in our case. For larger time scales quasi-static elastic moduli $\lambda = 4.88$~GPa and $\mu = 1.54$~GPa were used. To define viscoelastic properties, the following equally spaced relaxation times in a logarithmic scale were considered: $\tau_s = 0.01$, 0.03, 0.1, 0.3, 1, 3, 10, 30, 100 \textmu s. 
The ratio between $\xi_s$ and $\eta_s$ was chosen to be equal to $\lambda/\mu$ since the Poisson's ratio $\tilde{\nu}(\omega) = \tilde{\lambda}(\omega)/(2\tilde{\lambda}(\omega) + 2\tilde{\mu}(\omega))$ almost does not depend on frequency for high frequencies in glassy polymers~\cite{Tschoegl2002}. In this case the loss tangent is the same for all elastic moduli: $\tan \delta (\omega) = \lambda''(\omega)/\lambda'(\omega) = \mu''(\omega)/\mu'(\omega) $. For polystyrene the loss tangent is weakly dependent on frequency in very wide range of frequencies and approximately equals to 0.02 -- 0.03 \cite{Hurley2013, Benbow1958, Yadav2020}. To simulate this property, viscosity moduli were chosen as ${\xi_s = \xi}$ and ${\eta_s = \eta}$ for all relaxation times $\tau_s$. For relaxation times $\tau_s$, which are equally spaced in a logarithmic scale, one can estimate the loss tangent as
\begin{equation}
    \tan \delta \approx \frac{\pi N_{10}}{2 \ln10}\frac{\xi}{\lambda},
\end{equation}
where $N_{10}=2$ is the number of relaxation times per decade. To obtain the experimental value 0.03 of the loss tangent, given the values of $\lambda$ and $\mu$, the viscosity moduli were chosen as ${\xi = 0.09}$~GPa and ${\eta = 0.028}$~GPa. After all the assumptions made the quasistatic elastic moduli $\lambda$ and $\mu$ were the only tunable parameters. Their values, given earlier, were chosen to obtain good agreement between numerical simulations and experimental results.


The three-dimensional simulation was performed for the rectangular bar with a length of 200~mm and thickness of 10~mm with material properties mentioned above. The three-dimensional numerical analysis was performed using the multidomain pseudospectral method~\cite{Canuto2007-1, Canuto2007-2}. To understand the main effects of fast formation of the long wave, the numerical calculation was done without nonlinear effects. Reaching a balance between dispersion and nonlinearity for the formed long wave required longer computation time while the analytical properties of strain solitary waves have been previously studied in sufficient detail.

At the initial time moment the bar is unperturbed and a short normal stress is applied at its edge at $x = 0$:  
\begin{equation}
    F_x(y,z,t) = A_f e^{-t^2/\tau^2_{e}}, \quad F_y = F_z = 0,
\end{equation}
where $A_f = 1.4$~MPa and $\tau_e = 0.5\ \mu\text{s}$, while the other sides are free of external forces. In the experiment, the excitation pulse has much shorter duration. However, the shortening of the excitation pulse increases the simulation time significantly but leads to a minor change of the result. The value of $A_f$ was chosen to fit the amplitude of the formed long wave observed in the experiment.

\begin{figure}
	\centering
	\includegraphics[width=.68\linewidth]{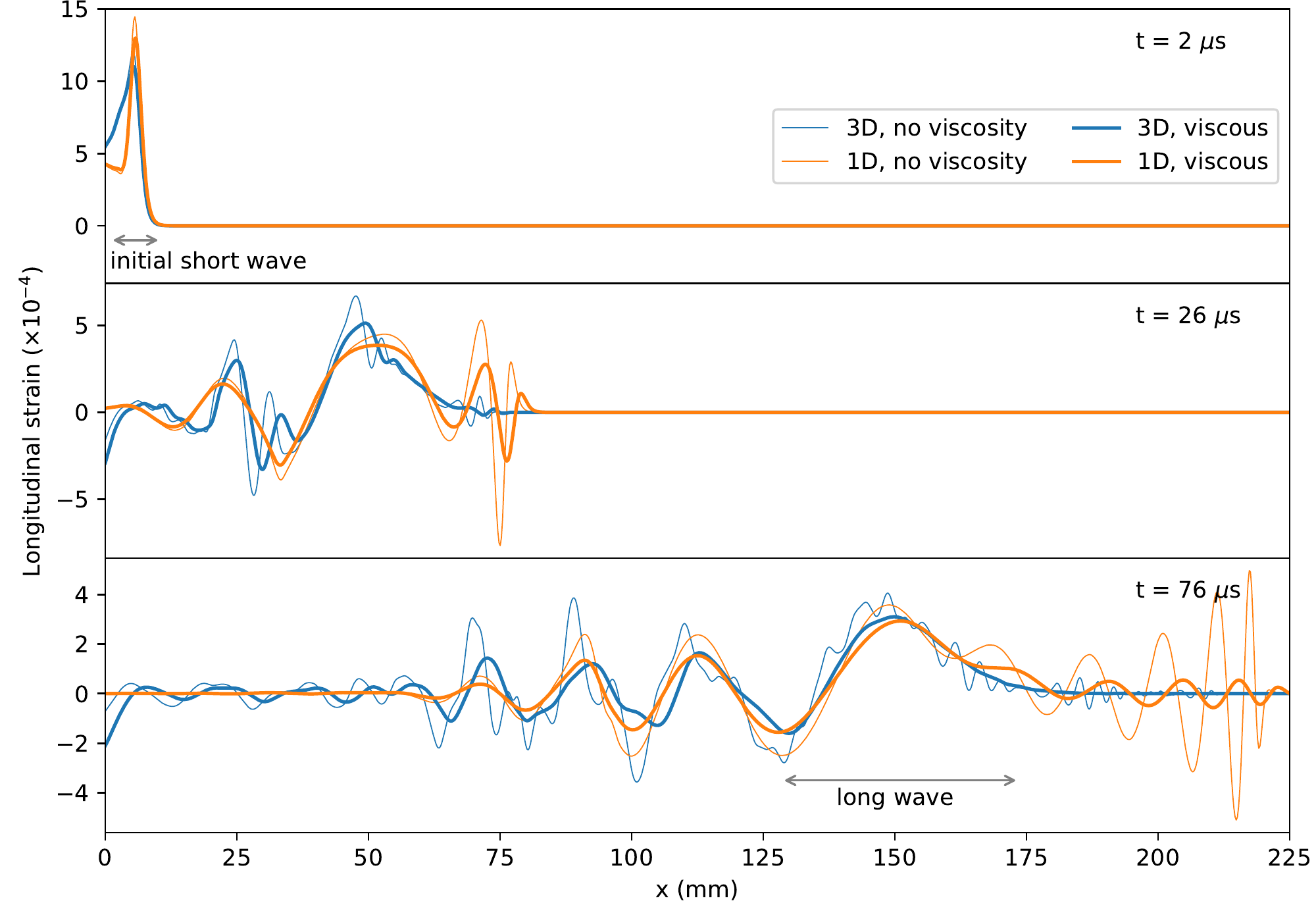}
    \caption{Longitudinal strain averaged over the cross-section as a function of the coordinate $x$ along the bar. Comparison of the full three-dimensional model and the simplified one-dimensional model with and without viscosity.}
    \label{fig:4curves}
\end{figure}

The results of numerical simulation obtained using the full three-dimensional model and simplified one-dimensional model are presented by thick lines in Fig.~\ref{fig:4curves}. In both cases one can see the formation of the leading long wave from the initial short wave. The position, the amplitude, and the width of the leading long wave are approximately the same in both models. However, the one-dimensional model shows some small oscillations in the front of the main wave and a slightly different form of the tail behind the main wave.

\begin{figure}
	\centering
	\includegraphics[width=\linewidth]{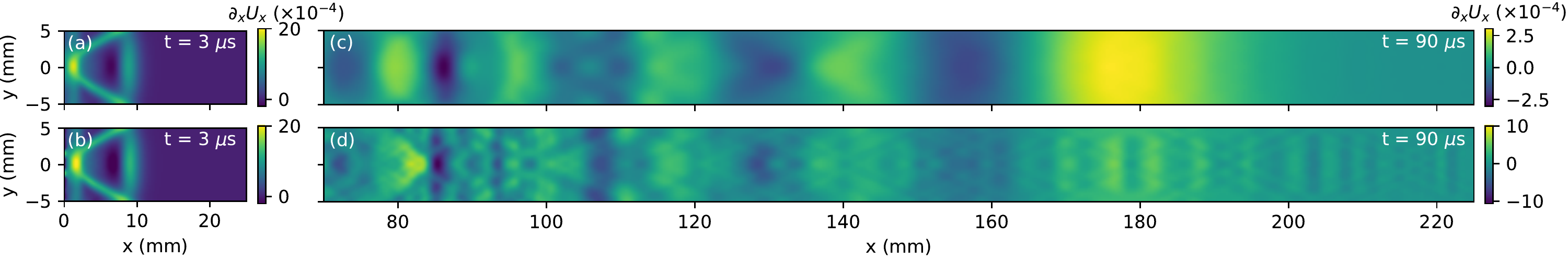}
    \caption{Results of the full three-dimensional simulation of the bar with generalized Maxwell viscosity (a, c) and without viscosity (b,~d). The panels show the longitudinal strain $\pd{x}U_x$ at the early and the late times as a function of $x$ and $y$ at $z=0$.}
    \label{fig:sim3d}
\end{figure}

To understand the effect of viscoelastic damping, we made the same calculations for an absolutely elastic case with $\xi_s=\eta_s=0$. According to Eq.~\eqref{lam_mu_complex}, the storage moduli $\lambda'(\omega)$ and $\mu'(\omega)$ have the frequency-dependent contribution introduced by $\xi_s$ and $\eta_s$. To compensate this effect, the storage moduli $\lambda'(\omega)=5.07$ GPa and $\mu'(\omega)=1.6$ GPa at a frequency of 0.1 MHz (i.e. for a typical duration of 10 \textmu s) were considered as elastic moduli $\lambda$ and $\mu$ for non-viscous case. The obtained results are presented by thin lines in Fig.~\ref{fig:4curves}. One can see approximately the same main long wave as was obtained by viscoelastic simulation. However, there are a number of short-wave oscillations, which were damped in the viscoelastic case. The viscoelastic effects are especially important for the full three-dimensional distribution of the strain presented in Fig.~\ref{fig:sim3d}. In the non-viscous case, formation of the long wave is completely hidden by a number of short waves.

\section{Discussion and conclusions}

\begin{figure}
	\centering
	\includegraphics[width=14cm]{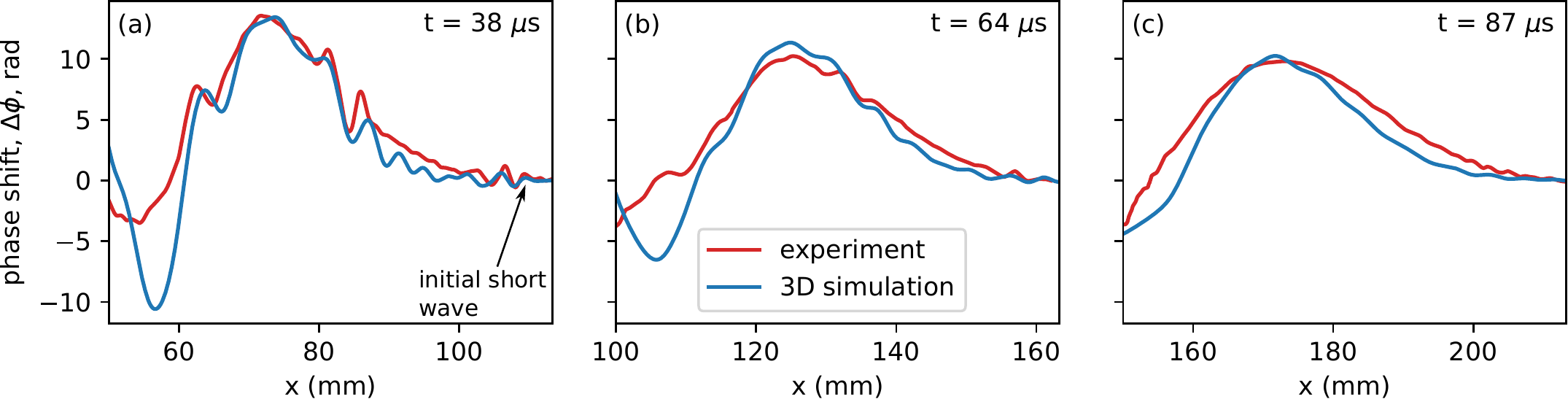}
	\caption{Comparison of the experimentally obtained phase shifts of the strain wave in the polystyrene bar with the results of three-dimensional numerical modelling at three delay times corresponding to the second, third, and fourth areas of the waveguide.}
	\label{fig:comp}
\end{figure}

In this paper we have detected formation of the long solitary wave using holographic technique. To explain the physical processes involved in the solitary wave formation we have performed numerical modelling of a polystyrene bar with realistic viscoelastic parameters. The comparison between the experimentally recorded phase shift and the phase shift obtained by three-dimensional numerical simulation is shown in Fig.~\ref{fig:comp}. One can see  approximately the same structure of the phase shift obtained at different time moments after the initial pulse. 
 
Therefore, the three-dimensional numerical model takes into account all necessary properties of the elastic waveguide, which are responsible for formation of the leading long strain wave. Since the three-dimensional model can be reduced to a simple one-dimensional model without a significant loss of precision, one can conclude that wave dispersion in the waveguide and viscoelastic properties of material lead to the formation of the long wave. The effects of dispersion can be understood as a reflection of the propagating elastic wave from sidewalls of the waveguide and concentration of the elastic energy. At the same time, viscoelastic effects suppress short-wave oscillations and reveal the formation of the leading long wave.

The further propagation of the long wave can be understood in terms of the solitary dynamics, which was discussed in \cite{Samsonov2001,TP2008,jap2010,jap2012,APL2014} and other papers referred therein. Recent results show that the nonlinear moduli $l$, $m$, $n$ may rapidly increase in their absolute values with decreasing frequency~\cite{Belashov2021}. Therefore, the influence of nonlinear properties may be much greater for the formed long solitary wave than for the initial short waves. 
However, short waves decay rapidly and nonlinear effects do not play a significant role in their evolution, which allows one to use frequency-independent nonlinear moduli with values at the characteristic frequency of a solitary wave.
Nonlinear viscoelastic dynamics with frequency-dependent moduli $l$, $m$, $n$ is the subject of the further study.

\section{Declaration of competing interests}
The authors declare that they have no known competing financial interests or personal relationships that could have appeared to influence the work reported in this paper.

\section{Acknowledgements}
The financial support from Russian Science Foundation under the grant \#17-72-20201 is gratefully acknowledged.

\appendix

\section{Appendix}
\label{app}
Nonlinear terms of the one-dimensional equation of motion:
\begin{subequations}
\begin{align}
    f_{\rm nl} =& \ 2\left(\hat{\lambda}+2l\right)(\pd{x}u)v + \left(\frac32\hat{\lambda}+3\hat{\mu}+l+2m\right)(\pd{x}u)^2 + \left(\hat{\lambda}+4l-2m+n\right)v^2 \nonumber\\
    &\  + \frac{R_*^{\smash{2}}}{2}\left(\hat{\lambda}+2\hat{\mu}+m\right)(\pd{x}v)^2, \\
    g_{\rm nl} =& \frac{1}{2}\left(\hat{\lambda}+2l\right)(\pd{x}u)^2  +\left(3\hat{\lambda}+3\hat{\mu}+4l+2m\right)v^2 +\left(\hat{\lambda}+4l-2m+n\right)(\pd{x}u)v \nonumber\\
    &\  - \frac{R_*^{\smash{2}}}{2}\left(\hat{\lambda}+2\hat{\mu}+m\right)\pd{x}[(\pd{x}u)(\pd{x}v)]   - \frac{R_*^{\smash{2}}}{2}\left(\hat{\lambda}+\hat{\mu}+m-\frac{n}{4}\right)\left(\pd{xx}(v)^2 - (\pd{x}v)^2\right),
\end{align}
\end{subequations}
where we keep only second-order terms and neglect higher order terms.

\bibliographystyle{apsrev4-1}  

\bibliography{refs.bib}


\end{document}